# Performance and Scalability Models for a Hypergrowth e-Commerce Web Site

Neil J. Gunther

*Performance Dynamics Company, 4061 East Castro Valley Boulevard,
Suite 110, Castro Valley, CA 94552, U. S. A.*

`njgunther@perfdynamics.com`
`http://www.perfdynamics.com/`

**Abstract.** The performance of successful Web-based e-commerce services has all the allure of a roller-coaster ride: accelerated fiscal growth combined with the ever-present danger of running out of server capacity. This chapter presents a case study based on the author's own capacity planning engagement with one of the hottest e-commerce Web sites in the world. Several spreadsheet techniques are presented for forecasting both short-term and long-term trends in the consumption of server capacity. Two new performance metrics are introduced for site planning and procurement: the *effective demand*, and the *doubling period*.

## 1      Introduction

Most Web sites are configured as a highly networked collection of UNIX and NT servers. The culture that belongs to these mid-range environments typically has not embraced the concepts of performance engineering in applications development. Moreover, many of the larger commercial web sites are large because they are success disasters. In other words, a suggested dot com business idea was implemented in a somewhat ad hoc fashion and later found to be far more attractive than site architects had originally anticipated. Now the site is faced with explosive levels of online traffic and this has brought capacity planning concerns to the forefront. Highly publicized Web sites such as Amazon.com, AOL.com, eBay.com, eToys.com, and Yahoo.com are just a few examples of success stories that have experienced variants of this effect.

Here, *success* is measured in terms of a site's ability to attract internet traffic. This is not necessarily equivalent to financial success. As many e-commerce Web sites have discovered (perhaps to the chagrin of Wall Street watchers), connections per second and dollars per connection are not always positively correlated. Nonetheless, it has become abundantly clear that server performance and site scalability are critical issues that can seriously affect the commercial viability of a Web site. The piece of the puzzle that is still missing is the concept of *planning* for site capacity. In fact, given the chaotic environments and undisciplined culture that are so typical of hyper-growth Web sites, one could be forgiven for thinking that *capacity planning* is a web-oxymoron [1]. There are, however, some encouraging signs that this situation is changing. A quick review of recent trade literature [2], [3], [4], [5] indicates that hyper-growth Web sites are beginning to appreciate the importance of application performance engineering and server scalability planning [6]. Even though additional server capacity is usually foreseen to be necessary, it must be procured well in advance of when it is actually needed. Otherwise, by the time the new servers arrive, traffic will have grown explosively and the additional capacity will be immediately absorbed with no advantage gained from the significant fiscal outlay.

In this chapter, we present a case study in capacity planning techniques developed by the author for one of the world's hottest web sites (which shall remain nameless for reasons of confidentiality). A notable attribute of the work described here is that the performance models used for planning must be lightweight and flexible to match the fast-paced growth of these environments [1]. The chapter falls logically into two parts. The first part (section 6) deals with determining short-term effective demand. The second part (section 7) uses peak effective demand statistics to project long-term server growth. Long-term growth projections, expressed in terms of the capacity *doubling time,* set the pace for the hardware procurement schedule.

## 2    Analysis of Daily Traffic

One of the most striking features of e-commerce traffic is the unusual variance in its intensity during the day. Mainframe capacity planners are already familiar with the bimodal "10-2" behavior of conventional commercial workloads due to online user activity [7]. The 10-2 designation refers to the twin peaks in CPU utilization occurring around 10am in the morning and 2pm in the afternoon (local time) within each nine-to-five day shift. There are two peaks because of the lunchtime slump in system activity around midday.

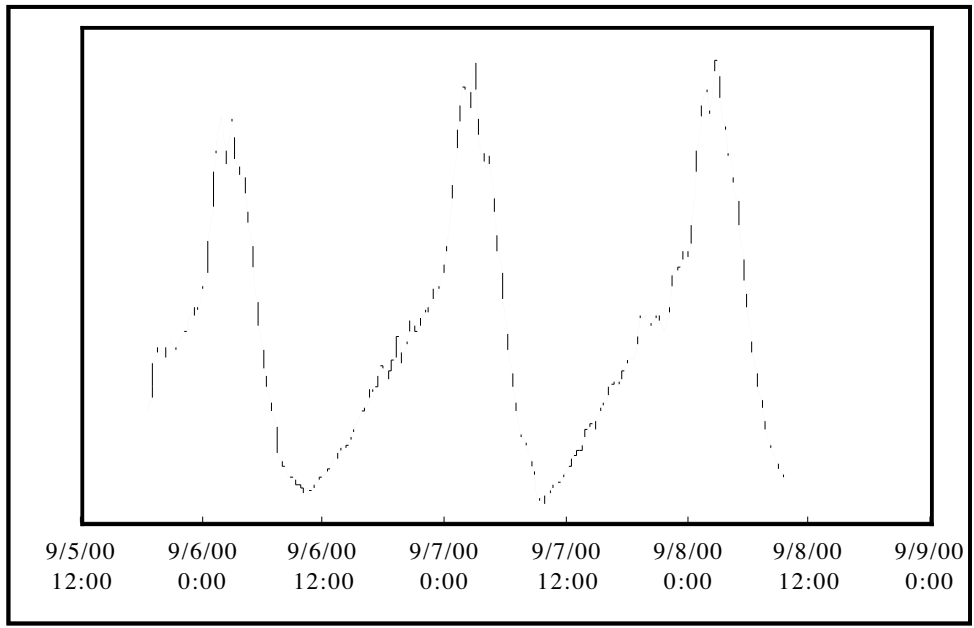

**Fig 1.** Characteristic daily peaks in traffic and resource utilization seen in many successful North American Web sites. Time is expressed as Coordinated Universal Time (UTC).

On the World Wide Web, however, the "day shift" can last for twenty-four hours and together with randomized traffic arrivals it is not clear whether any characteristic peaks exist at all; bimodal or otherwise. Worldwide access twenty-four hours a day, seven days a week, across all time zones suggests the traffic intensity might be very broad without any discernable peaks.

On the contrary! Not only did the measured traffic intensity peak, there was always a dominant peak around 2:00 hours UTC (Coordinated Universal Time). The single daily peaks in Fig. 1 also show up consistently in the measured CPU utilization of in-situ Web site servers as well as the measured outbound LAN (Local Area Network) packets inside the Web site. Moreover, this traffic characteristic is quite universal across all hyper-growth Web sites. The reason for its existence can be understood as follows.

**2.1 Bicoastal but Unimodal**

North American Web site traffic can be thought of as being comprised of two major contributions: one from the east coast and the other from the west coast. This naive partitioning corresponds to the two major population regions in the USA. If we further assume the traffic intensities (expressed as a percentage) are identical but otherwise phase-shifted by the three hours separating the respective time zones, Figure 2 shows how these two traffic profiles combine to produce the dominant peak. As the west coast contribution peaks at 100% around 4:00 hours UTC, the east coast contribution is already in rapid decline because of the later local time (7:00 hours actual UTC). Therefore, the aggregate traffic peak occurs at about 2:00 hours UTC.

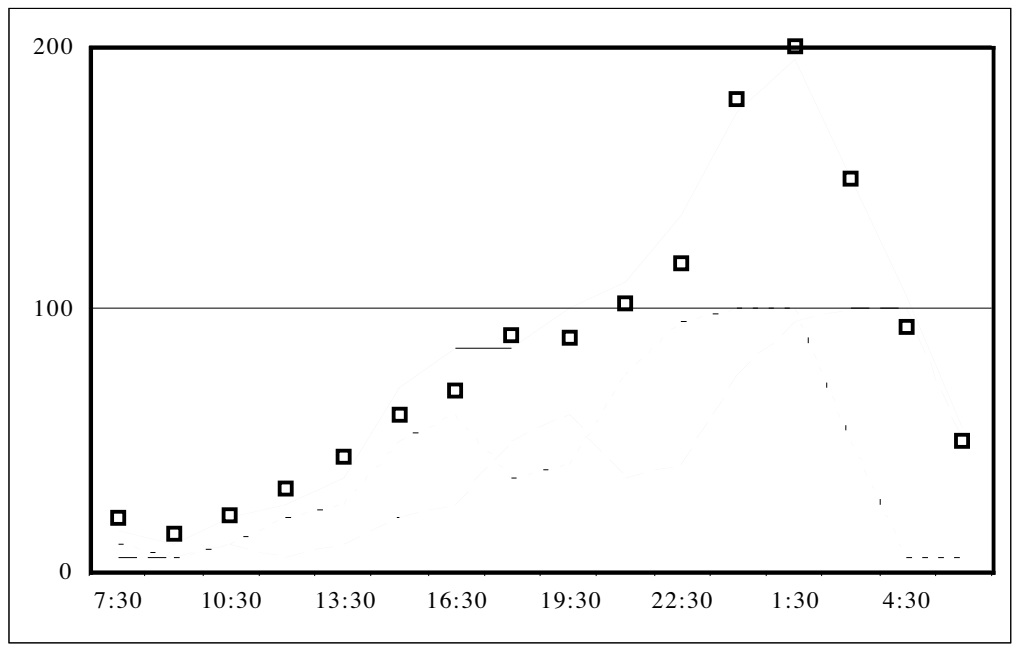

**Fig. 2.** Measured daily traffic intensity at a USA site (boxes). The single dominant peak (solid line) at 2am UTC is the sum of contributions from the east coast (dotted line) and the west coast (dashed line) phase-shifted by 3 hours for the respective time zones. As the western contribution is peaking, the eastern contribution is declining rapidly because of the later local time.

A benign bimodal character can be seen in the aggregate data of Fig. 2 with a small hump occurring around 16:30 hours UTC. This corresponds to a much more distinct bi-modality in both component traffic profiles and is reminiscent of the "10-2" mainframe traffic characteristic mentioned earllier but now reversed and shifted in time.

**2.2 All American!**

Another important conclusion can be drawn from this simple traffic analysis. There is essentially no significant traffic coming from time zones in Europe, Asia, Australia or the Pacific region. If these regions were contributing (albeit at less intensity) they would broaden the dominant peak significantly. Hence, the 10-2 bimodal profile seen in mainframe traffic can be understood as belonging to very localized usage where most users are connected to the mainframe by terminals or workstations and they are active in the same time zone. Although hyper-growth Web site traffic is delocalized into multiple time zones, it is still confined mostly to North America and within that continent, it is dominated by the activity of the coastal populations.

Several corollaries follow from these observations about the hyper-growth traffic profile:

- When it appears daily, we know we are looking primarily at North American users.

- This activity will be reflected in the consumption of other site resources e.g., server utilization, and network bandwidth.

- We can use it as a quick validation of any modeling predictions (e.g., section 6 and Fig. 3).

In the subsequent sections, we shall focus entirely on back-end server capacity rather than network capacity. The site in this study had plenty of OC12 bandwidth to handle outbound packets so, network capacity never caused any (non-pathological) performance problems.

Since the measured server utilization is bounded above by 100% at CPU saturation, a peak in server utilization caused by the characteristic traffic profile discussed above would remain obscured. One of the objectives in this study was to reproduce a server utilization profile that included such peaks. To this end we now introduce the concept of effective server demand.

## 3    Effective Server Demand

Effective demand is a measure of the work that is being serviced as well as the work that could be serviced if more capacity was available. It is similar to the mainframe notion of "latent demand" [7], [8]. The performance metric is dimensionless in the same way that CPU-busy is dimensionless but unlike CPU-busy, which is bounded above by 100%, effective demand can be expressed as an unbounded percentage. For example, an effective demand of 167% means that the application workload could have been serviced by one and two-thirds servers, even though only one completely saturated server was physically available to accommodate that workload.

Once a server becomes saturated, the run-queue necessarily begins to grow and this can have an adverse impact on user-perceived response time. Although there was some discussion about making response time measurements to gauge the expected impact of capacity upgrades, no commitment was undertaken by the site management during the engagement period.

### 3.1 Modeling Assumptions

The effective demand was calculated using the spreadsheet regression tool. The predictor is the effective demand (expressed as a percentage). Appropriate regressor variables can be determined from a factorial design analysis [9], [10], [11]. Typical regressors can include CPU clock frequency, the number of user submitting work to the system, the run-queue length, I/O rates, etc. One performance metric that cannot be used is the measured CPU utilization itself, since that is the variable we are trying to predict [8].

As with any statistical trend analysis of CPU capacity, we assumed that the CPU-intensive workload would remain so as more CPU capacity was added. It could happen that removing the CPU bottleneck by adding just a small amount of CPU capacity simply brings a disk or memory bottleneck to the surface and this would become the new scalability inhibitor.

### 3.2 Statistical Approach

There is no way to see such shifting bottlenecks a priori without the aid of more abstract performance tools; such as multi-class queueing models [12]. But a queueing model require that the underlying abstraction of the system be accurate. Without more sophisticated measurements than were available and more time (a precious commodity in hyper-growth Web sites) to validate such abstractions, the fallback to statistical data modeling was declared appropriate [13].

We analyzed ten weeks of performance data comprising two-minute samples of approximately two hundred metrics. During that period, many upgrades had been performed on the back-end servers of interest. These included upgrading CPUs and cache sizes, upgrading the version of ORACLE, and modifying the proprietary application software across the site. As a consequence, only a subset of that data was stable enough to employ for forecasting server capacity consumption.

## 4. Choosing Statistical Tools

As mentioned in section 3.1, it was decided to undertake a purely statistical analysis of the raw performance measurements of hyper-growth traffic at this Web site. A statistical analysis of this type typically uses analysis of variance (ANOVA [11]), regression [9], and time-series analysis (ARMIA [14]). However, because of the relatively short time over which performance metrics had been collected, no seasonal or other long-term periodic trends were present and time-series analysis was not needed.

Having elected to apply statistical analysis, the next issue was the selection of statistical tools to use for the analysis and, in particular, whether to purchase statistical software or build a custom application integrated into the current environment? The methods presented in this chapter had not been used elsewhere by the author and this meant that some prototyping was inevitable.

It was also not clear at the outset which statistical functions would be required. For example, implementing (which really means debugging) a set of statistical functions based on Numerical Recipes in C [15] or Perl [16] would be more time consuming than using an environment where a reasonably complete set of statistical functions already existed.

### 4.1 Spreadsheet Programming

The choice became obvious once it was realized that almost every desktop at the Web site had a PC running Microsoft Office and thereby had EXCEL readily available. What is not generally appreciated is that EXCEL is not just a spreadsheet [17]; it is a programming environment with Visual Basic for Applications (VBA) [18] as the programming language.

VBA is a reasonable prototyping language because it is object-oriented, comes with an integrated debugger, and has a macro-capture facility. As well as being ubiquitous, EXCEL is also platform-independent. The author could develop the tool off-site on an iMac and email it to the work place as an attachment to be downloaded on a resident PC for testing.

### 4.2 The Internet is Your Friend

Moreover, there are several Internet news groups [19] devoted to spreadsheets and VBA programming so that expert help is readily available for any obscure VBA programming problems. As we shall see later, data filtering was also an important requirement for selecting measured performance data according to some specified criteria. Plotting (or "charting" as it is called in EXCEL) is integrated into spreadsheets with a VBA programmable interface.

A broad set of standard statistical analysis functions (e.g., Regression, ANOVA, moving average) is available by default. This proved very useful because it was not known exactly which functions would be needed to analyze the performance data. EXCEL, as packaged, does not include the more sophisticated analysis tools (such as ARIMA) but these can often be found on the internet or as commercial add-in packages. Finally, there is an option to import data over the Web and also to publish spreadsheets as HTML pages.

## 5. Planning for Data Collection

Ideally, data center operations and capacity planning should be distinct groups. But for understaffed Web sites, economics often dictates the need for a leaner infrastructure than might otherwise be available traditionally.

Most data center operations, including hyper-growth Web sites, understand the need to collect performance data. Many data center managers are also quite prepared to spend several hundred thousand dollars on performance measurement tools. However, what is not yet well recognized in most operations centers is that the purpose for which such data is intended should determine how it is collected and stored. This means that operations center management has to think beyond immediate monitoring and schedule planning into a longer-term picture.

### 5.1 Use It or Lose It

The Web site in this study had purchased a commercial data collection product and had been using it for about six months prior to this capacity planning study. In order to conserve disk storage consumption these products have default settings whereby they aggregate collected performance metrics. For example, data might be sampled every few minutes but after eight hours all those sampled statistics (e.g., CPU utilization) will be averaged over the entire eight-hour period and thereby reduced to a single number. This is good for storage but bad for modeling and analysis.

Such averaged performance data is likely far too coarse for meaningful statistical analysis. Some commercial products offer separate databases for storing monitoring and modeling data.

Either the default aggregation boundaries should be reset, or operations management needs to be prescient enough to see that whatever modeling data is needed gets used prior to aggregation or is siphoned off and stored for later use. In either case, data collection and data modeling may need to be scheduled differently and well in advance of any data aggregation. This idea is new to many Web site data centers.

**5.2 Saved by the SymbEl**

And so it was with the site described here. Fortunately, alternative data was available because the platform vendor had installed their own non-commercial data collection tools written in the SymbEl or SE scripting language [20]. Since the user interface to this monitoring tool employed a java-enabled browser, and the job of the tool was simply to collect performance data in the background, it was called "SE Percolator". Thankfully, SE Percolator was unaware of sophisticated concepts such as aggregation and kept two-minute samples of more than two hundred performance metrics. It was this data that was used in our subsequent statistical modeling.

## 6. Short-term Capacity

As discussed in section 3.1, the effective CPU demand is calculated using regression techniques. We new present that technique in more detail.

**6.1 Multivariate Regression**

Because of a prior succession of changes that were made to the system configuration, only five weeks of data were stable enough to show consistent trending information. An EXCEL macro was used to analyze the raw metric samples and predict the effective demand using a multivariate linear regression model of the form

$$U^* = a_1 X_1 + a_2 X_2 + \ldots + a_0, \quad (1)$$

where each of the X's is a regressor variable, and the $\alpha$'s are the coefficients determined by an ANOVA analysis of the raw performance data. Here, $U^*$ refers to the estimated effective utilization of the server and it can exceed 100%.

**6.2 Spreadsheet Macros**

A Perl script [16] extracts the raw SymbEl Percolator data (section 5.2) averaging over 15-minute samples of the relevant time-stamped performance metrics for each day of interest. The extracted data is then read directly into a spreadsheet using the EXCEL Web Query facility. This produces about 100 rows of data in the spreadsheet. Two macros are then applied to this data.

The first of these filters out any row if it contains a measured CPU utilization of 95% or greater. Such rows are eliminated from the ANOVA calculations as being too biased for use by equation (1). The regressor variables (labeled $X_1$ through $X_6$ in Table 1) are then used by the macro to calculate the $\alpha$ coefficients of equation (1). Once these coefficients are known, the $U^*$ value

(third column in Table 3) is computed for each row in the spreadsheet. In general, the estimated and measured utilization will be fairly close until it gets near to saturation (e.g., 95% or greater), in which case significantly larger values of $U^*$ are estimated as shown in Figure 3. The complete VBA code for these modeling macros can be found in [1].

**Table 1.** Example of EXCEL spreadsheet layout for the multivariate regression model.

| DateTime | $U$ | $U^*$ | $X_1$ | $X_2$ | $X_3$ | $X_4$ | $X_5$ | $X_6$ |
|---|---|---|---|---|---|---|---|---|
| 9/29/99 0:00 | 25.25 | 32.90 | 32 | 19 | 16.45 | 18.96 | 15.04 | 131.56 |
| 9/29/99 0:16 | 27.25 | 36.85 | 45 | 11 | 17.01 | 22.49 | 14.18 | 136.08 |
| 9/29/99 0:32 | 47.12 | 54.01 | 50 | 42 | 29.52 | 33.32 | 27.07 | 236.13 |
| 9/29/99 0:48 | 45.88 | 51.19 | 53 | 38 | 27.29 | 32.09 | 24.62 | 218.29 |

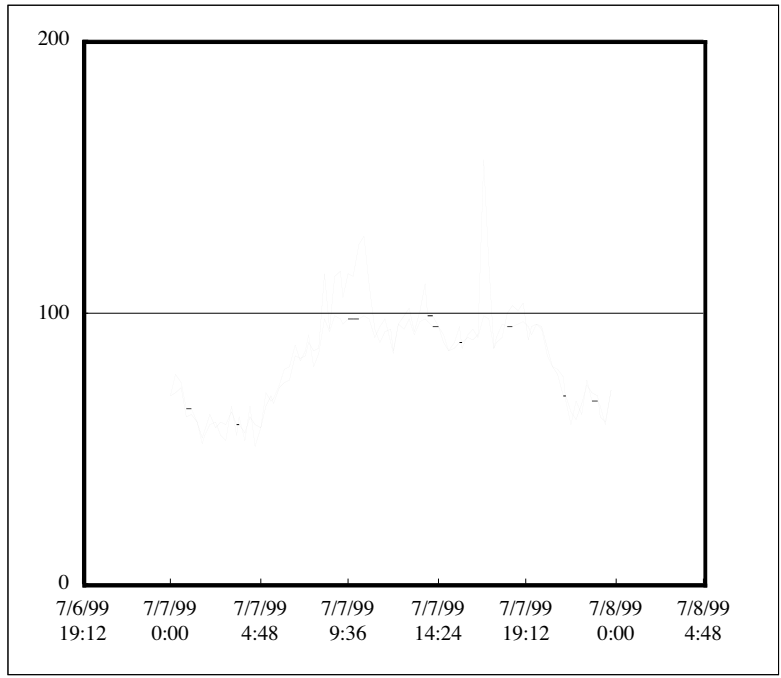

**Fig. 3.** Effective demand (solid) sometimes exceeding 100% compared with raw CPU utilization data (dashed line). Note the spike in utilization near 18:00 hours as discussed in section 2.

## 7. Long-term Capacity

Summary statistics from the short-term multivariate model described in Section 6 were then taken over into a weekly spreadsheet. About eight weeks worth of these summary statistics were needed to make reasonable long-term growth predictions.

### 7.1 Nonlinear Regression

For this phase of the exercise, a nonlinear regression model was also built in using spreadsheet macros. The maxima of the weekly effective demands, $U^*$, calculated from the multivariate model were fitted to a single parameter exponential model of the form,

$$U(w) = U_0 e^{bw}, \qquad (2)$$

where the long-term demand U(w) is expressed in terms of the numbers of weeks since the data analysis began. The constant $U_0$ is the CPU utilization at the zeroth week and the parameter 'b' is fitted by the Trending capabilities of EXCEL. That parameter specifies the growth rate of the exponential curve. For the data in Fig. 4, $U_0$ = 80.62 and b = 0.0309. We chose an exponential nonlinear server growth-model because it is expected that server capacity will need to scale beyond simple linear growth. Similar expectations exist for business growth models and it can be seen in the measured growth of network traffic.

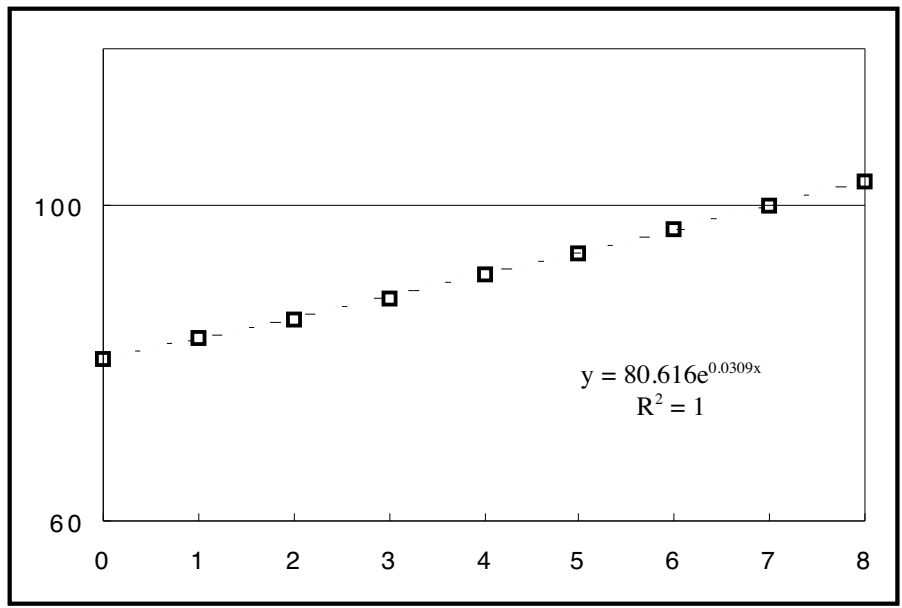

**Fig. 4.** Exponential regression (dashed line) on eight weeks of peak demand (boxes) from the short-term multivariate model in section 6. The curvature parameter b = 0.0309.

### 7.2 Doubling Period

A more intuitive grasp of the growth parameter comes from considering the time it takes to *double* the required server capacity. Thus, the natural logarithm of two divided by the rate parameter b in

Fig. 4 gives a doubling time corresponding to 5.6 months. In other words, every 6 months, twice as much server capacity will be consumed as is being consumed now!

This is at least an order of magnitude faster than the growth of typical mainframe data processing workloads and four times faster than Moore's Law [12] which predicts the number of transistors that can be packed into VLSI circuitry every two years. Nonetheless, this startling rate of growth agrees with growth of network traffic measured at this Web site

Fully loaded back-end database servers cost tens of millions of dollars. Considering that even the most lucrative hyper-growth Web sites only have revenue streams in the range of tens of millions of dollars per annum, a doubling time of six months amounts to nothing less than a Web definition of bankruptcy!

## 8. Upgrade Paths

The final task was to translate the results from these effective demand models into procurement requirements. This provides a set of upgrade curves corresponding to different possible CPU configurations, cache sizes, and clock speeds. Since the above regression models only pertain to the measurements on the current system configuration, we need a way to extrapolate to other possible CPU configurations.

### 8.1 Super-Serial Scaling

The next task was to translate this explosive growth rate into a set of sizing requirements for server procurement. For this purpose, we used the super-serial scaling model [12] which relates the effective capacity $C(p)$ of a server to the number of physical processors (p) as the function defined by,

$$C(p) = \frac{p}{1 + \sigma\{(p-1) + \lambda p(p-1)\}} \ . \qquad (3)$$

This function can be fitted to arbitrary measured CPU configurations. The throughput, $X(p)$, can then be predicted by multiplying an arbitrary measured throughput by $C(p)$ in equation (3) [22]. Fitting the super-serial model to server scaling data provides estimates for two parameters corresponding to such confounded hardware and software factors as:

- serial delays ($\sigma$) due to contention for mutex [20] and database locks [21].

- coherency delays ($\lambda$) due to stale cache refetching [21].

Because it would take us too far afield to present it here, the interested reader is encouraged to see references [12], [21] for a more extensive discussion of the super-serial model and its derivation.

The back-end servers at this web site were running the ORACLE database management system and could have up to 64-way CPUs. In the author's experience, the contention factor for ORACLE is typically around 3% or $\sigma = 0.030$ and the coherency term is typically $\lambda = 0.002$. This allowed us to make a realistic estimate of capacity factors for server configurations of interest.

Referring now to Table 2 which contains vendor scaling estimates, we see that adding 12 more 333 MHz CPUs with 4MB secondary cache to the existing 52-way, shows a 15% gain in headroom. But keeping the CPU configuration fixed at 52-way and increasing the CPU model from 333 MHz/4MB to 400 MHz/8MB, shows a 32% headroom gain. Performing both of these upgrades together shows a 52% gain in headroom. This seems rather optimistic for an ORACLE application.

**Table 2.** Vendor scaling data in transactions per minute. Reading left to right corresponds to changing CPU clock frequency ($\Delta$CLK) with the same CPU configuration, while reading downward corresponds to increasing the number of physical CPUs ($\Delta$CPU) at the same clock speed. These changes are also expressed as percentages in the outside column and row.

| System | 333/4 | 400/8 | $\Delta$CLK | Percentage |
|---|---|---|---|---|
| 52-way | 115,755 | 152,432 | 36,677 | 0.32 |
| 64-way | 133,629 | 175,969 | 42,340 | 0.32 |
| $\Delta$CPU | 17,874 | 23,537 | 60,214 | N/A |
| Percentage | 0.15 | 0.15 | N/A | 0.52 |

**Table 3.** Super-serial scaling model in transactions per minute. Reading left to right corresponds to changing CPU clock frequency ($\Delta$CLK) with the same CPU configuration, while reading downward corresponds to increasing the number of physical CPUs ($\Delta$CPU) at the same clock speed. These changes are also expressed as percentages in the outside column and row.

| System | 333/4 | 400/8 | $\Delta$CLK | Percentage |
|---|---|---|---|---|
| 52-way | 57,605 | 75,859 | 18,254 | 0.32 |
| 64-way | 60,875 | 80,165 | 19,290 | 0.32 |
| $\Delta$CPU | 3,270 | 4,306 | 22,560 | N/A |
| Percentage | 0.06 | 0.06 | N/A | 0.39 |

Referring to Table 3 which contains more conservative estimates, we see that adding 12 more CPUs to go from a 52-way to 64-way at the same clock speed 333 MHz and secondary cache size of 4MB, indicated a 6% gain in headroom. Keeping the CPU configuration fixed at 52-way but increasing the CPU clock speed from 333 MHz/4MB to 400 MHz/8MB, shows a 32% gain in headroom. Performing both of these upgrades together reveals a 39% headroom gain.

**8.2 Server Scalability Calculation**

Going from a 333 MHz to a 400 MHz clock while maintaining a 52-way configuration on the backplane and using the vendor scaling data in Table 2, we first calculate the increase ($\delta$) in CPU capacity from the tabulated throughput values:

$$\delta = \frac{X(400) - X(333)}{X(333)}$$
$$= \frac{36,677}{115,755}$$
$$= 31.7\%$$

Scalability envelopes corresponding to estimates in both Tables 2 and 3 and shown in Figure 5.

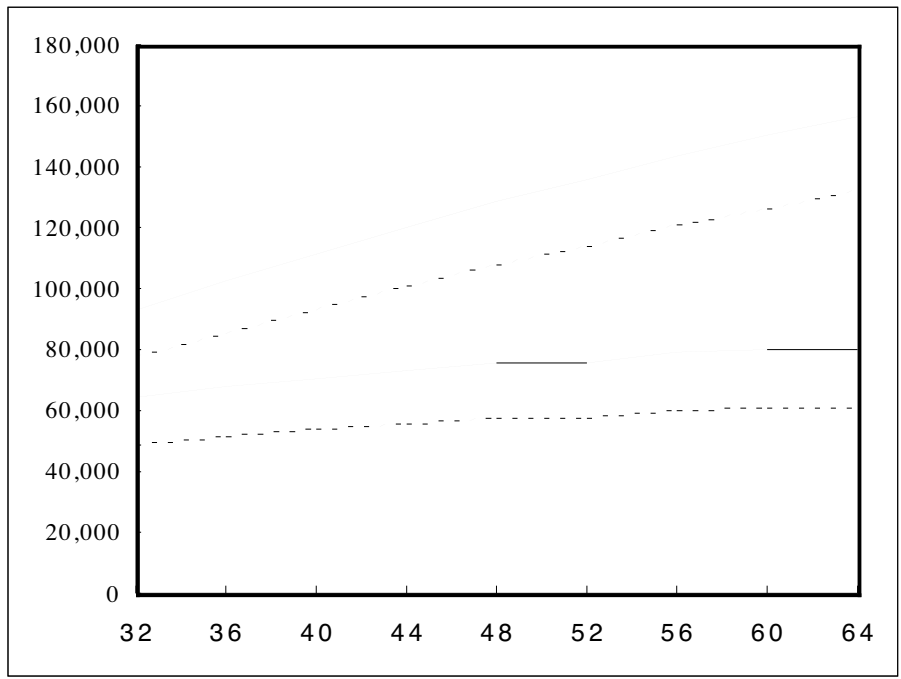

**Fig. 5.** Scalability envelopes corresponding to the values calculated in Tables 2 and 3. The solid lines represent the vendor (upper) and realistic (lower) CPU throughput (in transcactions per minute) for the 400 MHz/8 MB cache processor. Similarly, the dashed lines represent the vendor (upper) and realistic (lower) CPU throughput (in transcactions per minute) for the 333 MHz/4 MB cache processor.

Next, the increase in CPU capacity is used to adjust the weekly growth curve, C(w), downward because a lower curve takes longer to reach saturation. At week 20, $U_{333}(20)$ for the baseline server is 149.56 percent. By upgrading the CPU clock speed to 400 MHz, we know $\delta = 0.317$, and therefore the corresponding decrease in the effective demand can be determined as:

$$U_{400}(20) = U_{333}(20) \times (1 - 0.317)$$
$$= 149.56 \times 0.683$$
$$= 102.15$$

The corresponding long-term growth curves are shown in Figure 6 for the more conservative super-serial model.

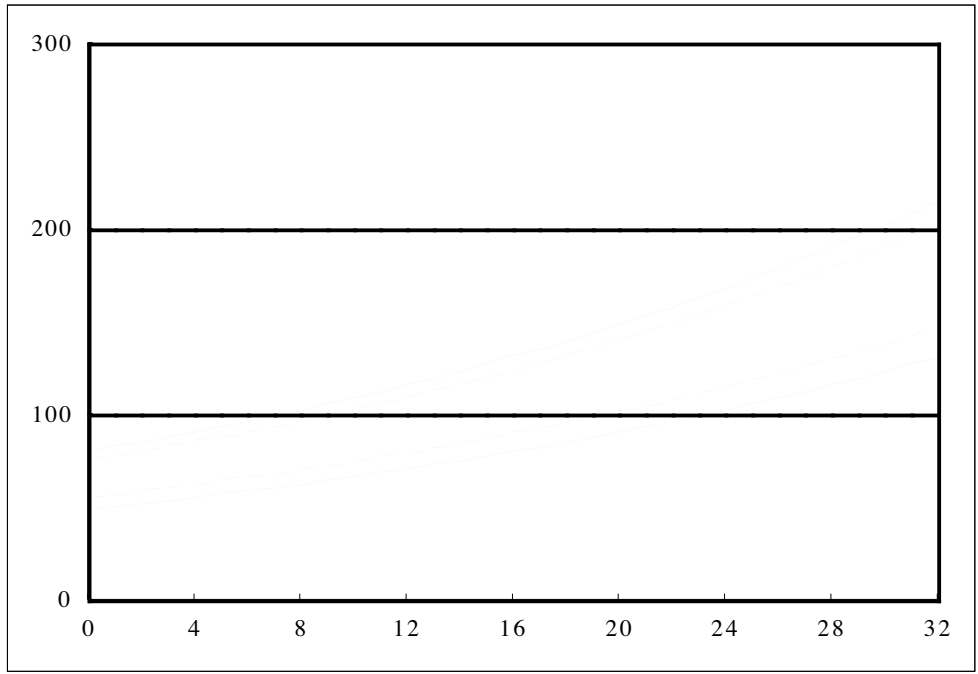

**Fig. 6.** Super-serial CPU capacity projections corresponding to the values calculated in Table 2. The x-axis represents the number of weeks since the capacity analysis data was begun. Week 20 corresponds to the Y2K boundary. The horizontal lines at 100 and 200 percent represent respectively a single server and two servers. The upper and lower solid curves represent respectively the worst case (maintain the current CPU configuration) and best case (fully loaded back-plane with fastest clock speed) projections.

Each of these scenarios could have been made more accurate if actual workload measurements had been available. Unfortunately, this Web site, like so many, was only in the early stages of using load test and software QA platforms prior to deployment. If the load testing had been further advanced, it might have been possible to use those workload measurements to improve some of our capacity predictions. In the meantime the site needed to make some fiscally responsible decisions regarding expensive server procurement and they needed to make those decisions quickly. Under those circumstance (not unusual, as we noted in the Introduction to this chapter) any sense of direction is more important that the actual compass bearing [13].

## 9. Summary

In this case study, we have shown how multivariate regression can used to analyze short-term effective CPU demand in highly variable data. The effective demand is a measure of the work that is being serviced as well as the work that could be serviced if more capacity was available. Daily data was collected and the *effective demand* calculated using spreadsheet macros. A weekly summarization of the peak effective demand was then used to generate a *nonlinear regression* model for long-term capacity growth and, in particular, extraction of *doubling* period.

The study presented in this chapter provided the first quantitative capacity model of the growth rate of this very successful Web site, A doubling period of about six months, was quickly recognized as having potentially devastating implications for the fiscal longevity of the Web site. Several performance engineering actions (that were not predicted by our capacity model) were immediately undertaken.

A purchase order for a clustered back-end server was written without further delay. In addition, the chief software developer immediately undertook to implement some thirty software changes that had been planned but never implemented. These software changes alone recovered thirty percent headroom on the existing back-end server and corresponded to increasing the doubling time to approximately 15 months. As a side effect, the model presented in this chapter was immediately invalidated!

Other important conclusions concerning hyper-growth Web sites also stemmed from our study. Almost all the significant traffic comes from just two North American time zones with very little coming time zones in Europe, Asia, Australia or the Pacific region. This relative localization of potential Web traffic is responsible for the large single peak discussed in section 2. The long-term growth of this peak is of prime concern for server procurement.

As Web connectedness continues to extend outside North America, this dominant peak can be expected to broaden and sustained capacity will become the new issue.

Most hyper-growth Web sites understand the need to collect performance data. Data center managers are prepared to spend hundreds of thousand of dollars on performance monitoring tools. What is not yet well recognized is that the purpose for which such data is intended should determine how it is collected and stored. This means that operations management has to think beyond immediate performance monitoring and schedule capacity planning collection and storage into a longer-term operations plan.